\documentclass[10pt]{article}
\usepackage{amsmath}
\usepackage{graphicx}
\usepackage{amsfonts}
\usepackage{amssymb}
\usepackage{epsf}
\usepackage{latexsym}

\def\80{\hspace{0.8in}}

\def\brho{\mbox{\boldmath$\rho$}}
\newcommand{\be}{\begin{enumerate}}
\newcommand{\ee}{\end{enumerate}}
\newcommand{\bi}{\begin{itemize}}
\newcommand{\ei}{\end{itemize}}
\newcommand{\bd}{\begin{description}}
\newcommand{\ed}{\end{description}}

\def\beq{\begin{equation}}
\def\eeq{\end{equation}}
\def\bea{\begin{eqnarray}}
\def\eea{\end{eqnarray}}
\def\pa{\partial}
\def\d{\textrm{d}}

\def\ttL{\mbox{\tt L}}
\def\ttD{\mbox{\tt D}}

\def\ttR{\mbox{\tt R}}

\def\cr{\mbox{\scriptsize{\bf $\mbox{ } \times \mbox{ }$}}}

\def\ma{\mbox{a}}

\def\mj{\mbox{j}}

\def\mn{\mbox{n}}

\def\mr{\mbox{r}}

\def\muu{\mbox{u}}

\def\mE{\mbox{E}}

\def\mL{\mbox{L}}

\def\mN{\mbox{N}} 

\def\mP{\mbox{P}}

\def\mR{\mbox{R}}

\def\sa{\mbox{\scriptsize a}}
\def\sb{\mbox{\scriptsize b}}

\def\sd{\mbox{\scriptsize d}}
\def\se{\mbox{\scriptsize e}}

\def\si{\mbox{\scriptsize i}}
\def\sj{\mbox{\scriptsize j}} 

\def\sll{\mbox{\scriptsize l}}  
\def\sn{\mbox{\scriptsize n}} 
\def\so{\mbox{\scriptsize o}} 
\def\sp{\mbox{\scriptsize p}}

\def\sr{\mbox{\scriptsize r}}
\def\sss{\mbox{\scriptsize s}}
\def\st{\mbox{\scriptsize t}}

\def\sA{\mbox{\scriptsize A}}

\def\sE{\mbox{\scriptsize E}}

\def\sH{\mbox{\scriptsize H}}

\def\sN{\mbox{\scriptsize N}}

\def\sR{\mbox{\scriptsize R}}

\def\sT{\mbox{\scriptsize T}}

\def\eph(B){\mbox{\scriptsize emergent(LMB)}}

\def\ta{\mbox{\tiny a}}

\def\bR{\mbox{{\bf R}}}

\def\bq{\mbox{\bf q}}
\def\br{\mbox{\bf r}}

\def\fE{\mbox{\sffamily E}}

\def\fP{\mbox{\sffamily P}}
\def\fQ{\mbox{\sffamily Q}}
\def\fR{\mbox{\sffamily R}}
\def\fS{\mbox{\sffamily S}}
\def\fT{\mbox{\sffamily T}}

\def\fV{\mbox{\sffamily V}}

\def\bn{\mbox{\bf n}}

\begin{document}
\begin{titlepage}
\vspace{.7in}
\begin{center}
 
\vspace{2in} 

{\LARGE\bf SHAPE SPACE METHODS FOR}

\vspace{.1in}
 
{\LARGE\bf QUANTUM COSMOLOGICAL TRIANGLELAND} 

\vspace{.2in} 

{\bf Edward Anderson}$^{1}$ 

\vspace{.2in}

{\em $^1$ DAMTP Cambridge U.K.} and 

{\sl Departamento de F\'{\i}sica Te\'{o}rica, Universidad Autonoma de Madrid.}  

\vspace{.2in}

\end{center}

\begin{abstract}

With toy modelling of conceptual aspects of quantum cosmology and the problem of time in quantum 
gravity in mind, I study the classical and quantum dynamics of the pure-shape (i.e. scale-free) 
triangle formed by 3 particles in 2-$d$.   
I do so by importing techniques to the triangle model from the corresponding 4 particles in 1-$d$ 
model, using the fact that both have 2-spheres for shape spaces, though the latter has a trivial 
realization whilst the former has a more involved Hopf (or Dragt) type realization.  
I furthermore interpret the ensuing Dragt-type coordinates as shape quantities: a measure of 
anisoscelesness, the ellipticity of the base and apex's moments of inertia, and a quantity proportional 
to the area of the triangle.  
I promote these quantities at the quantum level to operators whose expectation and spread are then 
useful in understanding the quantum states of the system.   
Additionally, I tessellate the 2-sphere by its physical interpretation as the shape space of triangles, 
and then use this as a back-cloth from which to read off the interpretation of dynamical trajectories, 
potentials and wavefunctions.  
I include applications to timeless approaches to the problem of time and to the role of uniform states 
in quantum cosmological modelling.    

\end{abstract}

\mbox{ } 

\vspace{0.7in}

\noindent PACS: 04.60Kz.

\mbox{ } 

\noindent{\bf Keywords}: Relational particle mechanics, toy models of Geometrodynamics, 
Problem of Time \& Quantum Cosmology.

\vspace{0.2in}

\noindent$^1$ ea212@cam.ac.uk \mbox{ } 

\mbox{ }

\end{titlepage}

\section{Introduction}

There have been various recent papers about small explicit examples \cite{AF, MGM, 08I, 08II, FORD, 
TriCl, SemiclI} of {\it relational particle mechanics (RPM)} models \cite{BB82, B03, ERPM, EOT, 
06I, 06II, FORD, 08I, Cones} for use as toy models for approaches to quantization schemes,  
\cite{SRGryb, 06I, 08II, Banal}, Quantum Cosmology \cite{BS89, SemiclI, 08II, AF, MGM, Forth} 
and approaches to the Problem of Time in Quantum Gravity \cite{BS89, K92, ParisBF08, 06II, SemiclI, 
Records, 08II, KieferSmolin08APOT, AF, MGM, ScaleQM, 08III, Forth}.
RPM's are mechanics models with no absolute notions of time, position, orientation, and, in the case of 
the present paper, scale, so that one has a dynamics of pure shape. 
Such characteristics motivate RPM's as whole-universe models for e.g. the Problem of Time in Quantum 
Gravity \cite{K81, K91K99, K92, I93, Rovelli, KieferSmolin08APOT} and other foundational issues in 
Quantum Cosmology \cite{HallHaw, EOT, H03, Rovelli}. 
RPM's are, in more detail, relational in Barbour's sense rather than Rovelli's (c.f. \cite{Rovelli, EOT, 
fqxi} and the comparisons in \cite{08I}.)    
Barbour-type relationalism consists of 1) {\it temporal relationalism} i.e. there is no 
meaningful primary notion of time for the universe as a whole.  
This is always 
implemented via reparametrization-invariant Jacobi-type actions \cite{Lanczos} at the classical level.   
2) {\it Configurational relationalism}: that a certain group $G$ of transformations are physically 
meaningless. 
This was originally implemented by Barbour--Bertotti \cite{BB82} and Barbour \cite{B03} (for scaled and 
{\it pure-shape}, i.e. scalefree, RPM's respectively) by the indirect means of including 
$G$-corrections to the velocities in the Jacobi-type action, variation then producing constraints 
which subsequently ensure that physical meaningless.
Furthermore, General Relativity (GR) -- in its guise as the geometrodynamics of 3-spaces that are 
compact without boundary -- fits this mold, since it can be cast in terms of a Jacobi-type action with 
the spatial 3-diffeomorphisms playing the role of $G$ \cite{BSWRWRABVanLanThanPhanLan2FEPI}.
RPM's can, moreover, at least in low dimensions, implement configurational relationalism instead by 
being directly constructed to be $G$-invariant \cite{FORD}; this approach builds on work of Kendall 
\cite{Kendall} on the spaces of shapes and then considers the `cones' \cite{Cones} over these 
so as to include scale.

As further motivation for RPM's, their usefulness as models of Quantum Cosmology and of the Problem of 
Time in Quantum Gravity stem from a number of analogies between them and geometrodynamics.  
As well as the above mention of their parallel with GR at the level of relational actions, there are 
parallels at the level of configuration space (see Sec 2.2) and various parallels which minisuperspace 
(a historically more studied toy model) \cite{MiniMagicHH83} does not (nontrivially) 
possess.\footnote{Minisuperspace is, however, closer to GR in having more specific and GR-inherited 
potentials and indefinite kinetic terms, so that it and RPM's are to some extent complementary in their 
similarities to GR, and thus in the ways in which they are useful as toy models of GR. 
Midisuperspace \cite{Midi} has both sets of features at once, but at the expense of Problem of Time 
calculations then often being intractably hard.}
%
Namely, 
1) an important feature of GR is its linear momentum constraint, which causes a number of substantial 
complications e.g. in attempted resolutions of the Problem of Time \cite{K92, I93}.  
RPM's zero total angular momentum constraint is a nontrivial analogue of this.  
2) GR has a notion of localization and thus of structure formation, which is lost in 
minisuperspace by treating all points as the same.  
RPM's, however, possess an analogous notion of `particle clumping'.  
Further analogies at the level of strategies for the Problem of Time are that RPM possesses an energy 
constraints that parallels GR's Hamiltonian constraint in producing frozen equations at the quantum 
level.  
Scaled RPM's also possess an Euler time \cite{06II, SemiclI} that closely parallels GR's York time 
\cite{York7273}, and a semiclassical approach \cite{SemiclI, MGM, ScaleQM, 08III, Forth} in close 
parallel with \cite{HallHaw} for GR.  
RPM's also possess counterparts of timeless \cite{NSI, Records, 08III} and histories theory 
\cite{Hartle, Forth} approaches to the Problem of Time. 
Further foundational issues in Quantum Cosmology that are addressable at least qualitatively using 
RPM toy models include whether structure formation in the universe have a quantum-mechanical origin 
\cite{HallHaw, SemiclI, Forth}, whether uniform states are important in Quantum Cosmology 
\cite{Penrose, Forth}, the problem of observables \cite{Rovelli}, dealing with QM for closed systems 
\cite{DeWitt, H03, 06I}, whether Quantum Cosmology is robust \cite{KR89} and the origin of the arrow of 
time \cite{EOT}.

Among these recent papers on RPM's, \cite{AF} concerns {\it scalefree 4-stop metroland}: a 
whole-universe model consisting of 4 particles on a line with no absolute scale, while \cite{08I, 08II} 
consider {\it scaled} and {\it pure-shape triangleland} models: whole-universe models consisting of the 
triangle made by 3 particles in the plane.  
Both 4-stop metroland and triangleland involve $\mathbb{S}^2$ reduced configuration spaces.  
Moreover (Sec 2), these spheres are realized distinctly in each case, with 4-stop metroland's being 
a simpler Cartesian realization, whilst triangleland's can be understood in terms of the Hopf 
map/Dragt-type coordinates \cite{Dragt} familiar from Molecular physics.
Unfortunately, 4-stop metroland \cite{AF} was studied after triangleland \cite{08I, 08II}, so that 
various techniques useful for simplification and physical interpretation became apparent only during 
the later and simpler 4-stop metroland study.
Moreover, I then found that these techniques have (interpretationally distinct) counterparts for the 
triangleland case's harder realization of the sphere, for which they again provide substantial 
simplification and physical interpretation. 
By these means, I write the current paper as a substantial improvement of the original triangleland 
papers \cite{08I, 08II} due to these subsequently found techniques and insights.   
The methods in question are 

\noindent A) tessellation of the mass-weighted shape space sphere by the physical interpretation 
(Sec 3).  
 This can then be used as a back-cloth from which one can read off which sorts of triangles are being 
picked out by classical trajectories as paths upon this (Sec 5), and by potentials and wavefunctions 
as height functions over this (Sec 6). 
(Moreover, addressing a wider range of physically meaningful questions than in \cite{08I} requires a 
wider range of orbital bases than in \cite{08II}, which I also provide for the first time in Sec 6.)  
 
\noindent B) Consideration of the `Dragt' shape quantities for triangleland's shape sphere (Sec 4).  
I provide a clear geometrical interpretation of these as an ellipicity of the system's moments of 
inertia, a measure of anisoscelesness and a quantity proportional to the area of the triangle (the last 
of which has yet further significance due to being coordinate/clustering independent).   
These are promoted at the quantum level to shape operators and then expectations and spreads for these 
are investigated (Sec 8).   
Sec 5 additionally comments on the physical interpretation of triangleland's conserved quantities. 

\noindent C) Use of the {\it na\"{\i}ve Schr\"{o}dinger interpretation} \cite{NSI,K92,I93} (Sec 9).    
This is one of the timeless approaches to the Problem of Time, by which only questions about the 
universe `being', rather than `becoming', a certain way, can be addressed.   
While this does have some practical limitations, it does address some questions of interest.
E.g., `what is the probability that the universe is almost uniform?   
One obtains answers to such questions by considering integrals of $|\Psi|^2$ over suitable regions of 
the configuration space, for $\Psi$ the wavefunction of the (model) universe.

\section{Notation and configuration spaces for RPM's}

A redundant configuration space for $N$ particles in dimension $d$ is $\fQ(N, d) = \mathbb{R}^{N d}$, 
coordinatized e.g. by particle position coordinates\footnote{I take  
$a$, $b$, $c$ as particle label indices running from 1 to $N$ (= 3 for triangleland)
$e$, $f$, $g$ as particle label indices running from 1 to $n = N - 1$ for relative position variables. 
$i$, $j$, $k$ as spatial indices.
$p$, $q$, $r$ as relational space indices running from 1 to 3.
$u$, $v$, $w$ as shape space indices running from 1 to 2. 
Furthermore, presentation is enhanced by denoting particular coordinate components with {\sl downstairs} 
indices.} 
$\bq^a$.  
Rendering absolute position meaningless by passing to any sort of relative coordinates leaves one on 
{\it relative space} = $\fR(n, d) = \mathbb{R}^{n d}$, most conveniently coordinatized by {\it relative 
Jacobi coordinates} \cite{Marchal} ${\bR}^e$ (combinations of relative positions $\br^{ab} = \bq^b - 
\bq^a$ between particles and inter-particle clusters such that the kinetic term is diagonal, with 
associated cluster masses $\mu_e$).  
In fact, I use {\sl mass-weighted} relative Jacobi coordinates ${\brho}^e = \sqrt{\mu_e}\bR^e$ (see Fig 
\ref{Fig1}), their squares the partial moments of inertia $I^e = \mu_e|{{\bR}^e}|^2$ and the normalized 
versions of the former, $\bn^e = \brho^e/\rho$, where $I$ is the total moment of inertia and $\rho := 
\sqrt{I}$ is the hyperradius.  
I label all of these coordinates according to the clustering information which each requires for 
specific use as follows.  
I use \{a...c\} to denote a cluster composed of particles a, ... c, ordered left to right in 1-$d$ and 
anticlockwise in 2-$d$. 
I take these to be distinct from their right to left and clockwise counterparts respectively, i.e., 
I study unoriented configurations (the mathematics for which is simpler, as already pointed out in 
\cite{Aqui86}).
I insert commas to indicate a clustering, i.e. a partition into clusters. 
These notations also cover collisions, in which constituent clusters collapse to a point.  
In triangleland, I use (a) as shorthand for \{a, bc\} where a,b,c form a cycle.  
In 4-stop metroland, I use (Hb) as shorthand for \{1b,cd\} (H-shaped clustering) where b,c,d form a 
cycle.  


{            \begin{figure}[ht]
\centering
\includegraphics[width=0.6\textwidth]{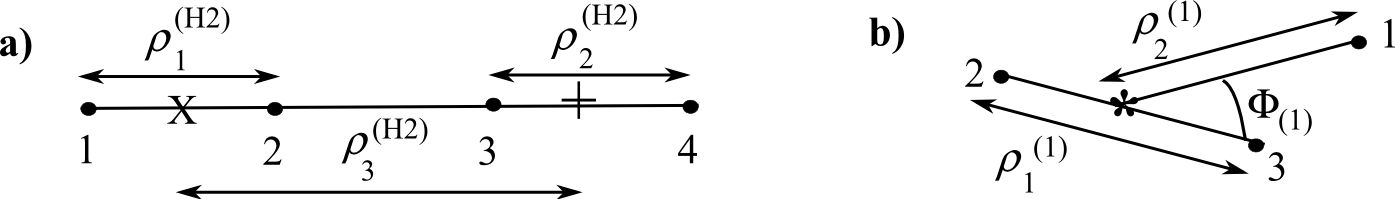}
\caption[Text der im Bilderverzeichnis auftaucht]{        \footnotesize{ 
X, + and * are the centres of mass of clusters \{12\}, \{34\} and \{23\} [cluster \{12\} consists of 
particles 1 and 2 etc.]

\noindent
a) I indicate one permutation of mass-weighted relative Jacobi H-shaped coordinates for 4-stop 
metroland.

\noindent
b) I indicate one permutation of relative Jacobi coordinates for triangleland. 
In {\sl mass-weighted space}, the magnitudes are the length of a base and what is a median in the  
equal-mass case that this paper focusses on for simplicity.    
I furthermore define $\Phi_{(\sa)} = \mbox{arccos}\big( \brho_1^{(\sa)}\cdot\brho_2^{(\sa)} /
\rho_1^{(\sa)}\rho_2^{(\sa)} \big)$ as the `Swiss army knife' angle between the two $\brho^e_{(\sa)}$. 
     }         }
\label{Fig1}\end{figure}          }

If, instead, rotation with respect to absolute axes is to have no meaning, then one is left on 
{\it relational space} ${\cal R}(n, d) = \fR(n, d)/\mbox{Rot}(d)$ (for Rot($d$) the $d$-dimensional 
rotation group, which is the $G$ for scaled RPM).
If absolute scale is also to have no meaning, then one is left on \cite{Kendall} {\it preshape space} = 
$\fP(n, d) = \fR(n, d)/$Dil (for Dil the dilational group);  it is straightforward to see this is 
$\mathbb{S}^{nd - 1}$.   
If both of the above are to have no meaning, then one is left on \cite{Kendall} {\it shape space} = 
$\fS(n, d) = \fR(n, d)/\mbox{Sim}(d)$ for Sim$(d) = \mbox{Rot}(d) \times \mbox{Dil}$ the 
$d$-dimensional similarity group, which is the $G$ for pure-shape RPM).  
For $d$ = 1, eliminating the rotations is trivial, so that $\fS(n, 1) = \fP(n, 1) = \mathbb{S}^{n - 1}$; 
this gives that 4-stop metroland's shape space is $\mathbb{S}^2$.
For $d$ = 2, $\fS(n, 2) = \fR(n, 2)/SO(2) \times \mbox{Dil} = \mathbb{S}^{2n - 1}/U(1) = 
\mathbb{CP}^{n - 1}$ \cite{Kendall, FORD}, while the well-known result $\mathbb{CP}^{1} = \mathbb{S}^2$ 
then gives that triangleland's shape space $\fS(3, 2) = \mathbb{S}^2$ also.    
Finally, relational space ${\cal R}(n, d)$ can then be envisaged in shape--scale split form as the 
{\it cone} $C(\fS(n, d))$ over the corresponding shape space $\fS(n, d)$ \cite{Cones}.

The above suite of configuration spaces is analogous to GR's under the following correspondence, 
which adds motivation to the study of the RPM program.    
The configuration space Riem($\Sigma$) of positive-definite metrics over a fixed topology $\Sigma$ 
(taken for simplicity to be a compact without boundary one) corresponds to $\fR$($n$,$d$), the 
3-diffeomorphism group Diff($\Sigma$) corresponds to Rot($d$) and the subsequent quotient space 
superspace($\Sigma$) = Riem($\Sigma$)/Diff($\Sigma$) \cite{DeWittFischer} to ${\cal R}$($n$,$d$). 
The conformal transformations Conf($\Sigma$) correspond to Dil, and the subsequent configuration space 
\cite{FM96} pointwise superspace($\Sigma$) = Riem($\Sigma$)/Conf($\Sigma$). 
Quotienting both out yields \cite{York7273, York74ABFO, FM96} conformal superspace, CS($\Sigma$) = 
Riem($\Sigma$)/Diff($\Sigma$) $\times$ Conf($\Sigma$), analogous to RPM's shape space $\fS$($n$,$d$).   
Finally, \{CS + V\}($\Sigma$), where the V is global volume, is the closest thing known to a `space of 
true dynamical degrees of freedom for GR' \cite{York7273, ABFKO} is analogous to analogous to the 
shape-scale split form of ${\cal R}(n,d)$. 
(CS and CS + V are well-known from the initial value problem literature, and are also connected 
to York's internal time candidate for GR \cite{York7273, K81, K92, I93}, for which scaled RPM's 
also possess an analogue, Euler time \cite{06II, SemiclI}.)

Furthermore, whether by direct formulation \cite{FORD} or upon reduction from indirect formulation 
\cite{06II, TriCl, FORD, 08I, AF}, 4-stop metroland and triangleland both have the standard spherical 
kinetic metric. 
I study all of these and the sometimes useful analogy of the 2-sphere in ordinary space under the 
`umbrella notation' of ($\alpha$, $\chi$) as azimuthal and polar spherical angles respectively. 
I then denote the triangleland case of this by ($\Theta$, $\Phi$) [where $\Theta_{(\sa)} = 2\,
\mbox{arctan}({\rho}_1^{(\sa)}/\rho_2^{(\sa)})$], the 4-stop metroland case by ($\theta$, $\phi$) 
[where $\theta_{(\sH\sb)} = \mbox{arctan}\left(\sqrt{\rho_1^{(\sH\sb)\,2}+ \rho_2^{(\sH\sb)\,2}}
/\rho_3^{(\sH\sb)}\right)$, $\phi_{(\sH\sb)} = \mbox{arctan}\left(\rho_2^{(\sH\sb)}/\rho_1^{(\sH\sb)}
\right)$ and the 2-sphere is ordinary space case by ($\theta_{\sss\sp}$, $\phi_{\sss\sp}$).  
Then the kinetic term for mechanics on whichever of these spheres is 
\beq
\fT = (\dot{\alpha}^2 + \mbox{sin}^2\alpha\,\dot{\chi}^2)/2 \mbox{ } .   
\label{Kin}
\eeq

\noindent
It is furthermore mathematically and quantum-mechanically useful to recast the above `umbrella 
treatment' in terms of three variables $\muu^{\Delta}$ such that $\sum_{\Delta = 1}^3\muu^{\Delta \, 2} 
= 1$, i.e. as unit Cartesian vector components for some Euclidean 3-space taken to surround the sphere,  
$\muu_x = \mbox{sin}\,\alpha\,\mbox{cos}\,\chi$, 
$\muu_y = \mbox{sin}\,\alpha\,\mbox{sin}\,\chi$,   
$\muu_z = \mbox{cos}\,\alpha$.

\section{Tessellation of triangleland's shape sphere by its physical interpretation}

\cite{08I, 08II} followed one clustering \{1,23\} algebraically via a set of coordinate systems 
corresponding to it [denoted with (1) subscripts].     
While this picks out the \{23\} double collision [D(1) point] and  the (1)-merger [M(1) point: 
configuration for which particle 1 is at the centre of mass of \{23\}], the \{12\} and \{13\} 
counterparts of these points are hard to see in that representation.
Labelled triangles also have A) three distinct notions of isoscelesness (one per choice of clustering 
\{a, bc\} with bc as the base bisected by the perpendicular from the apex a). 
Each of these corresponds to a bimeridian on the triangleland shape sphere that separates hemispheres 
of left- and right-slanting triangles.
B) three distinct notions of regularness (equal partial moments of inertia for the base pair and for 
the apex, or, equivalently, median = base in the case of equal masses). 
Each of these corresponds to a bimeridian on the triangleland shape sphere that separates hemispheres 
of sharp (median $>$ base) and flat (median $<$ base) triangles.  
Then \cite{08I, 08II}'s (1)-clustering picks out (1)-isoscelesness, (1)-leftness, (1)-rightness, 
(1)-regularness, (1)-sharpness and (1)-flatness well, but the corresponding (2)- and (3)- notions are,  
likewise, hard to see.  
This Section's graphical approach enables one to see simultaneously and clearly the (1)-, (2)- and 
(3)-clusterings' versions of all these things (Figure 2).
Then classical trajectories can be interpreted as paths upon this tessellated sphere, and classical 
potentials and quantum-mechanical probability density functions as height functions over it.  
Triangleland also possesses some clustering-independent notions: A) collinear configurations, which 
are orientationless and correspond to the equator, which separates hemispheres of clockwise and 
anticlockwise oriented labelled triangles.
The D's and M's lie equally-spaced on this equator.
B) Equilateral triangle configurations of the two orientations, corresponding to the poles (denoted by 
E and $\bar{\mE}$).

{            \begin{figure}[ht]
\centering
\includegraphics[width=0.9\textwidth]{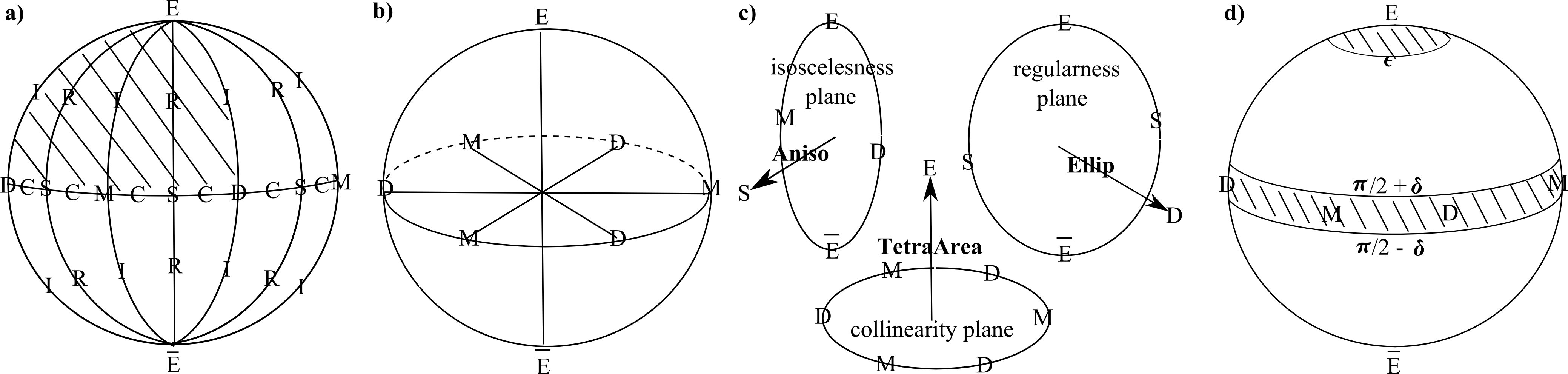}
\caption[Text der im Bilderverzeichnis auftaucht]{        \footnotesize{ a) The sphere, when 
interpreted as the triangleland shape sphere, is tessellated by 24 equal isosceles spherical triangle 
faces.  
This is also true of the 4-stop metroland shape sphere, but each case is geometrically different, 
triangleland's being in the shape of the 12 signs of the Zodiac further partitioned into boreal and 
austral skies, whilst 4-stop metroland's is in the shape of the cube plus dual octahaedron (see 
\cite{AF, Magnus}).
Triangleland's tessellation has 36 edges, 12 that are quarterings of the three bimeridians of 
isoscelesness (I), 12 that are quarterings of the three bimeridians of regularness (R), and the 
12-fold split of the equator of collinearity (C). 
Finally, triangleland's tessellation has 14 vertices: equilateral triangles E, 
$\bar{\mbox{\footnotesize E}}$ at the poles, with the other 12 vertices on the equator forming the 
sequence ``double collision (D), spurious (S), merged (M), spurious"... 
By ``spurious", I mean that they are relevant insofar as they are where bimeridians of regularness and 
the equator of collinearity intersect, but, unlike the D's, M's and E's, these points appear to be 
bereft of any {\sl extra} physical or geometrical significance. 
As regards mapping faces to faces, the corresponding group is the order-48 `austroboreal zodiac' group 
$\mathbb{D}_{12} \times \mathbb{Z}_2$.
However, upon taking into account the labellings of the vertices and edges by distinct physical 
meanings, this group is broken down to a smaller one: the order-12 trigonal bipyramidal group 
$\mathbb{D}_3 \times \mathbb{Z}_2 \mbox{ }$; this is reflected by the fundamental region of the  
{\sl labelled} tessellation being as shaded.  
The latter group is, moreover, isomorphic to $\mbox{ } S_3 \times \mathbb{Z}_2$, corresponding to  
the freedom of relabelling the 3 partices and of ascribing an overall orientation.     

\noindent b) Note that how the triangleland picture is represented in this paper has been rotated 
through $\pi/2$ relative to the presentation in \cite{08I}, to the more natural presentation with E 
and $\bar{\mbox{\footnotesize E}}$ on the poles, leaving all three D's on an equal footing on the 
equator, rather than the cluster-dependent picking out of one of the D's to be the North pole.
The D and M vertices then pick out 3 particularly distinguished axes at $2\pi/3$ to each other in the 
collinearity plane, in 1 to 1 correspondence with the 3 clusterings of Jacobi coordinates.  

\noindent c) The relation between Sec 4's basis of Cartesian vectors and the planes perpendicular 
to each, clearly illustrating that TetraArea is a departure from the plane of collinearity, 
that anisoscelesness Aniso is indeed a departure from the plane of isoscelesness, 
and that ellipticity Ellip is a departure from the plane of regularity. 
This is useful in the interpretation of the orbitals in Sec 6.  

\noindent d) Near-equilateral configurations correspond to spherical caps $\Theta \leq \epsilon$ and 
Near-collinear configurations correspond to the belt around the equator, $\pi/2 - \delta \leq \Theta 
\leq \pi/2 + \delta$. 
This is an example of mapping propositions about shape in space to propositions about regions 
of shape space; such mappings are useful e.g. in the study of timeless approaches (see e.g. Sec 8, 
v1 of this paper's preprint and \cite{AF}).}        }
\label{Fig-NEW-2}\end{figure}            }

\section{Shape coordinates and their physical interpretation}


I denote by [a] the natural bases with principal axis in the cluster-independent E direction, 
second axis in the D(a) direction and third axis in the perpendicular S(a) direction [Fig 2c)]; 
I denote spherical coordinates with respect to these by $(\Theta_{[\sa]}, \Phi_{[\sa]})$. 
I denote by (a) the cluster-centred bases which is like [a] but with principal and second axes 
reversed; spherical coordinates with respect to these are ($\Theta_{(\sa)}$,$\Phi_{(\sa)}$).  
These are the nicest basis as regards physical interpretation of conserved quantities, and are also 
adapted to a particularly simple combination of HO-like potentials that I term 
`special'.
To adapt to more complicated HO-like potentials,\footnote{Without loss of generality about the (1)-clustering [and 
dropping (1)-labels], the harmonic oscillator (HO)-like potential $\fV = \sum
\mbox{}_{\mbox{}_{\mbox{\tiny $I = 1$}}}^{2}K_{I}\sn^{2}_I/2 + L\,\bn^1\cdot\bn^2 = 
A + B\,\mbox{cos}\,\Theta + C\,\mbox{sin}\,\Theta\,\mbox{cos}\,\Phi$ for constants $K_i = H_i/\mu_i$ 
(Jacobi--Hooke coefficient over Jacobi cluster mass), $L$ a cross-term of similar nature, $A = (K_1 + 
K_2)/16$, $B = (K_1 - K_2)/16$ and $C = L/8$.
The $C = 0$ problem is then the `special' one, with conserved quantity ${\cal J}$ (see Sec 5.1).
I also term $B = C = 0$ problem `very special' (this amounts to a constant potential problem via 
the springs happening to balance out, and possesses 3 conserved quantities).
HO-like potentials are motivated by their being quantum-mechanically well behaved 
\cite{08II, AF}, reasonably analytically tractable and their scaled counterparts appearing in analogue 
models of Cosmology \cite{Cones}.}
one can also take E as principal axis while placing the 
2-axis at a general angle $\gamma$ within the collinearity plane (measured without loss of generality 
from the (1)-axis; I term this the $[\gamma]$-bases. 
Finally, one could again interchange the principal and second axis allocations; I term these the 
$(\gamma)$-bases.


The 4-stop metroland and triangleland $\mathbb{S}^2$ shape spaces have very different mathematical and 
physical interpretations.  
For 4-stop metroland, there is an obvious triple of Cartesian directions $\mn^i$ (relative separations 
between clusters) for the $\muu^{\Delta}$ to be.   
For triangleland, however, there are four and not three relative Jacobi vector components.

However, for triangleland relative separations have not three but four components, so their obvious 
relation is now to a $\mathbb{S}^3$, which can, however, be Hopf-mapped to the $\mathbb{S}^2$, which 
leaves a much less obvious set of 3 Dragt-type \cite{Dragt}\footnote{See 
\cite{GronwallSmith62} for earlier literature and e.g. \cite{MTIwai87PPAqui93LRML99} for some 
applications.
I say `Dragt-type' rather than `Dragt' due to some slight differences between 2-$d$ and 3-$d$, oriented 
and unoriented triangles, and the scaled and pure-shape cases.}
coordinates in the role of the $\muu^{\Delta}$:
\beq
\mbox{dra}^{(\sa)}_x                                       = 
2\,\bn_1^{(\sa)} \cdot \bn_2^{(\sa)}                         = 
2\mn^1_{(\sa)}\mn^2_{(\sa)}\,\mbox{cos}\,\Phi_{(\sa)}      = 
\mbox{sin}\,\Theta_{(\sa)}\,\mbox{cos}\,\Phi_{(\sa)}       = 
\mbox{sin}\,\Theta_{[\sa]}\,\mbox{cos}\,\Phi_{[\sa]}       =
\mbox{dra}^{[\sa]}_x
\mbox{ } ,
\label{dragt1}
\eeq
\beq
\mbox{dra}^{(\sa)}_y                                       = 
2|\bn_1^{(\sa)} \cr \bn_2^{(\sa)}|                         = 
2\mn^1_{(\sa)}\mn^2_{(\sa)}\,\mbox{sin}\,\Phi_{(\sa)}      =
\mbox{sin}\,\Theta_{(\sa)}\,\mbox{sin}\,\Phi_{(\sa)}       = 
\mbox{cos}\,\Theta_{[\sa]}                                 =
\mbox{dra}^{[\sa]}_z                                       
\mbox{ } ,
\label{dragt2}
\eeq
\beq
\mbox{dra}^{(\sa)}_z                                       =  
\mn_2^{(\sa)\, 2} - \mn_1^{(\sa)\, 2}                      =
\mbox{cos}\,\Theta_{(\sa)}                                 = 
\mbox{sin}\,\Theta_{[\sa]}\,\mbox{sin}\,\Phi_{[\sa]}       = 
\mbox{dra}^{[\sa]}_y
\mbox{ } .   
\label{dragt3}
\eeq
In these depending on squared quantities, one can see consequences of scaled triangleland having 
$\rho^2 = I$ and not $\rho$ as radial scale variable (which arises from coordinate range issues 
\cite{Cones}).
Thus the $\muu^{\Delta}$ representation of the sphere for 4-stop metroland goes to $\mn^e_{(\sH \sb)}$ 
via the Cartesian map, and, for triangleland to $\mbox{dra}^{\Delta}_{(\sa)}$ via the Hopf/Dragt map.


For 4-stop metroland \cite{AF}, the $\muu^{\Delta} = \mn^e$ are straightforwardly interpretable as 
relative sizes of the constituent clusters.  
For triangleland, the $\muu^{\Delta} = \mbox{dra}^{\Delta}$ have a more complicated but nevertheless 
highly geometrically meaningful interpretation, parts of which are new to the present paper and is also 
of interest for Molecular Physics.  
In the natural basis, $\mbox{dra}_z^{[a]}$ is a measure of departure from collinearity to clockwise and 
anticlockwise oriented triangles; it is moreover clustering-independent alias a democracy invariant 
\cite{LR95}.
It is furthermore $4 \times$ area (the area of the triangle per unit $I$ in mass-weighted space), 
so I denote it by TetraArea.      
$\mbox{dra}^{[\sa]}_y$  is a measure of departure from (a)-regularness toward (a)-sharp or (a)-flat 
triangles. 
It is moreover $\mn_2^{[\sa]\,2} - \mn_1^{[\sa]\,2}$ i.e. the `ellipticity' of the two `normalized' 
partial moments of inertia involved in the a-clustering, so I denote it by Ellip(a).  
dra$^{(\sa)}_x$ is a departure from (a)-isoscelesness toward (a)-left or (a)-right slanting triangles, 
so I denote it by Aniso(a) for `a-anisoscelesness', which it is specifically in the sense that aniso(a) 
$\times \mbox{ } I$ per unit base length in mass-weighted space is the $l_1 - l_2$ indicated in Fig 3, 
i.e. the amount by which the perpendicular to the base fails to bisect it.
N.B. that the Ellip, Area and Aniso combinations are endemic in \cite{08I, 08II}'s solutions, so that 
understanding and using these combinations 1) greatly simplifies the formulae involved. 
2) It brings out the parallels with \cite{AF} and the 2-sphere in space, permitting many checks of 
the triangleland solutions (while the problems remain analogous, the study of general HO-like 
potentials does eventually break the analogy).  

{            \begin{figure}[ht]
\centering
\includegraphics[width=0.2\textwidth]{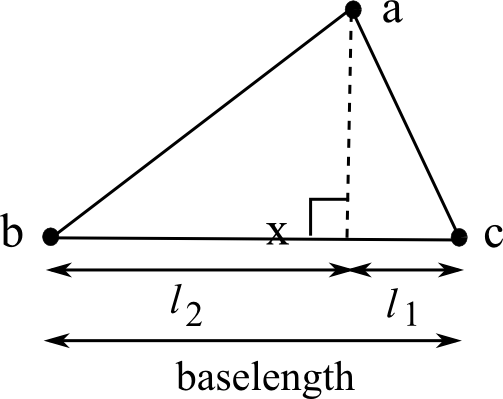}
\caption[Text der im Bilderverzeichnis auftaucht]{        \footnotesize{ Definitions of $l_1$ and $l_2$ in interpretation of this 
paper's notion of anisoscelesness.]                          }        }
\label{Fig-3tri}\end{figure}            }

\section{Classical formulation, solutions and interpretation}

The temporal relationalism implementing Jacobi action is $2\int \sqrt{\fT(\fE - \fV)}\d\lambda$.   
For triangleland, this is most conveniently represented for $\fT$ the `pure-shape kinetic term' given by 
(\ref{Kin}) with $\dot{\mbox{ }} := \d/\d\lambda$ for $\lambda$ a label time parameter, which is of 
physical dimension energy/$I$, $\fV$ 1/4 of the potential term and $\fE$ a constant playing a similar 
role to total energy in mechanics; these last two have the physical dimensions of energy $\times 
\mbox{ } I$.  
In this representation, the pure-shape theory consistency condition \cite{B03, 08I} takes its most 
natural form, namely that $\fE - \fV$ is required to be homogeneous of degree 0.
This formulation amounts to a direct implementation of configurational relationalism with respect to 
Rot($d$) $\times$ Dil for triangleland or Dil for 4-stop metroland. 
Its equations of motion are of the standard spherical form
\beq
\alpha^{{\prime\prime}} - \mbox{sin}\,\alpha\,\mbox{cos}\,\chi\, \chi^{{\prime}2} = 
- {\pa{V}}/{\pa\alpha} \mbox{ } \mbox{ } , \mbox{ } \mbox{ } 
(\mbox{sin}^2\alpha\, \chi^{{\prime}})^{{\prime}} = - {\pa{V}}/{\pa\chi}
\label{Su} \mbox{ } ,
\eeq

\noindent 
interpreting $\alpha$ as $\Theta$, $\chi$ as $\Phi$ and $V$ as $\fV$.
For RPM's, the time derivative $\prime$ is to be interpreted as $* := \sqrt{(\fE - \fV)/{{\fT}}}
\mbox{ } \dot{\mbox{ }}$ is the derivative with respect to an emergent time.  
The conjugate momenta are then $p_{\alpha} = \alpha^*$ and $p_{\chi} = \mbox{sin}^2\alpha\,\chi^*.$

\subsection{The physical nature of triangleland's conserved quantities}

Denote the three objects associated with the isometry group of the sphere, $SO(3)$, by 
\beq
\ttR_{\Delta} := \epsilon_{\Delta\Lambda\Gamma}\muu^{\Lambda}\muu^{\Gamma\,*}
\eeq
and let $\ttR_{\sT\so\st\sa\sll} = \sum_{\Delta}\ttR_{\Delta}^2$.  
It follows from (\ref{Su} ii) that one of the $\ttR_{\Delta}$ occurs as a conserved quantity if the 
potential is independent of $\chi$; all three occur if the potential is additionally independent of 
$\alpha$.  
For the sphere in `actual space' these conserved quantities are, physically, angular momenta 
$\ttL_{i}$ and $\ttL_{\sT\so\st\sa\sll}$.
For 4-stop metroland, they cannot physically be angular momenta since it is a spatially 1-d model; 
they are, rather \cite{AF}, {\it dilational momenta} ${\cal D}^{(\sH\sb)}_f = {\epsilon_{efg}}
\mn^f_{(\sH \sb)}\mn^{g}_{(\sH \sb)}\mbox{}^* = \ttD^{(\sH\sb)}_g
\mn_{(\sH \sb)}^f/\mn_{(\sH \sb)}^g - \ttD^{(\sH\sb)}_f\mn_{(\sH \sb)}^g/\mn_{(\sH \sb)}^f$, where 
$\ttD^{(\sH\sb)}_e = \rho^ep_e$ (no sum, $p_e$ conjugate to $\rho^e$) are partial dilations.
Dilational and angular momentum are subcases of a concept that generalizes angular momentum (see also 
\cite{Smith}), which \cite{AF} identify to be {\it rational momentum} (i.e. corresponding to a ratio 
which does not necessarily have an interpretation as an angle in physical space).  
Now, for triangleland, there are also three obvious individual $SO(3)$ objects -- one per clustering -- 
which are now relative angular momenta ${\cal J}_{(\sa)}$.   
However these three are coplanar (and thus not independent), and at $2\pi/3$ to each other and as such 
{\sl only one} of them can be chosen to pertain to a given orthogonal triplet of $SO(3)$ objects:     
${\cal R}^{(1)}_3 := {\cal J}^{(1)}_3$.  
The other two objects are then found to be linear combinations of a relative angular momentum and a 
relative dilational momentum.
This clarifies the physical meaning of the lengthy formulae in \cite{08I} for what we now denote 
by ${\cal R}_1$ and ${\cal R}_2$.\footnote{In \cite{08I}, they are denoted by $J_1$ and $J_2$,   
a notational difference which highlights two misconceptions that the current paper clarifies.  
Firstly, these were wrongly assumed to also be relative angular momenta (hence the letter $J$).  
Secondly, the scaled and pure-shape theories' conserved quantities looked to differ by a factor of $I$, 
and so were denoted by $J_{\Delta}$ and ${\cal J}_{\Delta}$ respectively.  
However, they are in fact the same once proper account is taken of the difference in physical dimension 
between each theory's $\fT$'s and $(\fE - \fV)$'s, which cancels out that factor of $I$.
Thus $J_{\Delta} = {\cal J}_{\Delta}$, the 1 and 2 components of which are physically 
${\cal R}_{\Delta}$, rather!}
%
Triangleland's total rational momentum is then ${\cal R}_{\sT\so\st\sa\sll} = \sum_{\Delta}
{\cal R}^{(1)}_{\Delta \, 2} = {{\cal D}_{(1)}}^2 + {{\cal J}_{(1)}}^2/\mbox{sin}^2\Theta_{(1)}$ 
[= 2$\fT$].

\subsection{Using this paper's techniques to describe a number of classical solutions}

The very special HO combination is constant over $\mathbb{S}^2$.
Its solutions are great circles, with the tessellation revealing that a number of these have very clear 
interpretations: the equator is a periodic motion through all the collinear configurations including all 
3 D's and all 3 M's. 
Some bimeridians go through all of a clustering's notions of isosceles including E, $\bar{\mE}$, M and D.
Some go through all of a clustering's notion of regular, corresponding to 1 particle orbiting the centre 
of mass of the other two so that $\Phi$ augments forever whilst $I_1$ and $I_2$ stay fixed at $I$/2 each.

The special HO potential is heart- or spheroidal-shaped, each with one end bulkier than the other, thus 
giving a well centred on, depending on the sign of $B$ (i.e. which of $K_1$ and $K_2$ is larger), 
either i) on the $D(1)$ point (thus having a propensity to trap classical trajectories inside a region 
of (1)-sharp triangles).
The physics here is that the inter-cluster spring binding \{23\} to the `external particle' 1 is 
stronger than \{23\}'s intra-cluster spring.
Or ii) on the $M(1)$ point [the (1)-flat triangle counterpart, with the physics here now being that 
the intra-cluster spring of \{23\} is more tightly binding than the inter-cluster spring between \{23\} 
and 1.] 
For ${\cal J} \neq 0$, in each case add narrow infinite skewers along the axis of symmetry, thus 
precluding the most very (1)-sharp and (1)-flat triangles.
Other HO's have the same heart or spheroidal shapes but centred on another of the equator of 
collinearity's points, $\mbox{arctan}(C/B) = \gamma$.   
E.g. $\gamma = \pi/2$ (pure $C$ term) picks out the regular triangles.  
Finally note that the triangleland potential breaks the tessellation group for ${\cal J} \neq 0$: 
the heart/spheroidal shape has symmetry group $\mathbb{D}_{\infty}$ and there is also a skewer along 
some axis perpendicular to the E-axis, so the overall problem retains just a $\mathbb{Z}_2$ 
reflection symmetry about the plane of collinearity.

\section{Quantum triangleland}

The kinematical quantization involves three quantities $\muu^{\Delta}$ such that $\sum_{\Delta = 1}^3  
\muu^{\Delta\,2} = 1$ and the three components of the $SO(3)$ rational momenta $\ttR_{\Delta}$.  
One can take these to be $x_i$ and $\ttL_i$  for the 2-sphere in space, $\mn^e_{(\sH\sb)}$ and 
${\cal D}^{(\sH\sb)}_e$ for 4-stop metroland, and, for triangleland, the $\mbox{dra}^{\Delta}_{(\sa)}$ 
[given by eq's (\ref{dragt1}-\ref{dragt3}) and interpreted as Aniso(a), TetraArea and Ellip(a) as 
per Sec 4, and nominally promoted to operators by hatting] 
and the ${\cal R}^{(\sa)}_{\Delta}$ of Sec 5.1.

The time-independent Schr\"{o}dinger equation for the HO-like potential problem on 
triangleland is then
\beq
\widehat{\cal R}_{\sT\so\st\sa\sll}^2\Psi = ({\cal A} + {\cal B}\,\widehat{\mbox{dra}}\mbox{}^{(1)}_z\ + 
{\cal C}\,\widehat{\mbox{dra}}\mbox{}^{(1)}_x)\Psi \mbox{ } .
\eeq
Thus, e.g. in (1)-centred spherical coordinates \cite{08II}, 
\beq 
(\mbox{sin}\,\Theta_{(1)})^{-1}(\mbox{sin}\,\Theta_{(1)}\Psi_{,\Theta_{(1)}})_{,\Theta_{(1)}} + 
(\mbox{sin}\,\Theta_{(1)})^{-2}\Psi_{,\Phi_{(1)}\Phi_{(1)}} =
({\cal A} + {\cal B}\,\mbox{cos}\,\Theta_{(1)} + 
{\cal C}\,\mbox{sin}\,\Theta_{(1)}\,\mbox{cos}\,\Phi_{(1)})\Psi  \mbox{ } .  
\label{spheTISE}
\eeq

\noindent 
Here, ${\cal A} = 2(A - \fE/4)/\hbar^2$, ${\cal B} = 2B/\hbar^2$ and ${\cal C} = 2C/\hbar^2$. 
The above uses the conformal ordering; for 2-$d$ configuration spaces as in the present paper, this 
furthermore coincides with the sometimes also-advocated  Laplacian ordering. 
(See \cite{Banal} and references therein for arguments for these operator orderings in quantum 
cosmological modelling.)

\subsection{The very special quantum problem on triangleland}

The very special case ${\cal B} = {\cal C} = 0$ then has a potential that balances out to be constant.  
Thus it is mathematically the same as the linear rigid rotor, for which the Hamiltonian is 
$\ttL_{\sT\so\st\sa\sll}$ up to multiplicative and additive constants, so, effectively, this and 
$\ttL_3$ form a complete set of commuting operators whose eigenvalues and eigenfunctions are the 
well-known spherical harmonics.
In our triangleland counterpart, however, our `rigid rotor' is in configuration space rather than in 
space and with total relative rational momentum $\widehat{{\cal R}_{\sT\so\st\sa\sll}} = 
\sum_{\Delta = 1}^3 \widehat{{\cal R}}^{(1)\,2}_{\Delta}$ in place of total angular momentum and 
projected relative rational momentum $\widehat{{\cal R}}^{(1)}_3$ in place of magnetic angular momentum.  
These then have eigenvalues $\hbar^2\mR(\mR + 1)$ and $\hbar\mr$ respectively, for R and r 
respectively the {\it total} and {\it projected `relative rational momentum quantum numbers'} (analogous 
to the ordinary central force problem/rigid rotor total and magnetic angular momentum quantum numbers, 
l and m respectively, and to the total and projected `relative dilational momentum quantum numbers' of 
4-stop metroland \cite{AF}).  
This is solved by
\beq
\Psi_{\sR\sr}(\Theta, \Phi) \propto Y_{\sR\sr}(\Theta, \Phi)
\propto \mP_{\sR}^{\sr}(\mbox{cos}\Theta)\mbox{exp}(\pm i\mr\Phi) \mbox{ } 
\label{YrR}
\eeq
for $\mP_{\sR}^{\sr}(Z)$ the associated Legendre functions in $Z$, while R $\in \mathbb{N}_0$ and r  
is such that $|\mr| \leq \mR$.  
Also, $\mR(\mR + 1) = - {\cal A}$, which here means that $\fE = 2\hbar^2\mR(\mR + 1) + K_2$ is 
required of the model universe's energy and inter-cluster effective spring in order for there to be 
any quantum solutions ($\fE$ is {\sl fixed} as this is a whole-universe model so there is nothing 
external from which it could gain or lose energy).  
If this holds, there are then 2R + 1 solutions labelled by r (one can see the preceding sentence 
cuts down on a given system's solution space, though the more usual larger solution space still exists 
in the `multiverse' sense, see the Conclusion for more on this).  
Moreover, (\ref{YrR}) has further value in also occuring as a separated-out part of the corresponding 
scaled relational particle model problem \cite{08III}.

I use the natural/equilateral [a] bases new to this paper and which are well-adapted to questions about 
equilateral and collinear configurations.   
The form of the solution is then\footnote{I use
\cite{AF}'s ${\cal T}_{\sr}(\xi) := T_{\sr}(\xi)$              for cosine solutions and 
                                $:= \sqrt{1 - T_{\sr}(\xi)^2}$ for sine solutions, where 
$T_{\sr}(\xi)$ is the Tchebychev polynomial of rth degree in some variable $\xi$ 
for the $\Phi$-part of the wavefunction.} 
\beq
\Psi_{\sR\sr}\left(\Theta_{[\sa]}, \Phi_{[\sa]}\right) \propto 
\mP_{\sR}^{\sr}(\mbox{TetraArea}){\cal T}_{\sr}
\left(
\mbox{Ellip(a)}/\sqrt{1 - \mbox{TetraArea}^2}
\right) \mbox{ } .
\eeq
Triagleland's orbitals in this basis then have the following physical interpretations. 
R = 0 = r is the s-orbital, so it favours all triangles equally. 
All r = 0 solutions give each clustering equal status by axisymmetry.  
R = 1, r = 0 is the orbital $p_{\sd\sr\sa_z}^{[\sa]} = p_{\sT\se\st\sr\sa\sA\sr\se\sa}$ so non-area i.e.  
collinearity is nodal and big area and so approximate equilaterality is favoured. 
The R = 1 = r sin orbital is $p_{\sd\sr\sa_y}^{[\sa]} = p_{\sE\sll\sll\si\sp}^{[\sa]}$ so non 
(a)-ellipticity i.e. (a)-regularity is nodal and big (a)-ellipticity i.e. (a)-sharp and (a)-flat 
triangles are favoured, and the cos orbital is $p_{\sd\sr\sa_x}^{[\sa]} = 
p_{\sA\sn\si\sss\so}^{[\sa]}$ so (a)-isoscelesness is nodal and big (a)-anisoscelesness i.e. (a)-right 
and (a)-left slanting triangles are favoured.  
R = 2, r = 0 is the orbital $d_{\sd\sr\sa_z^2}^{[\sa]} = d_{\sT\se\st\sr\sa\sA\sr\se\sa^2}$ so it has 
a collinearity ring and two approximaterly-E polar lobes.  
The R = 2, r = 1 sin orbital is $d_{\sd\sr\sa_y\sd\sr\sa_z}^{[\sa]} = 
d_{\sE\sll\sll\si\sp\,\sT\se\st\sr\sa\sA\sr\se\sa}^{[\sa]}$ so (a)-regularness and collinearity are 
nodal, with (a)-isoscelesness elsewise favoured, and the cos orbital is 
$d_{\sd\sr\sa_x\sd\sr\sa_z}^{[\sa]} = d_{\sA\sn\si\sss\so\,\sT\se\st\sr\sa\sA\sr\se\sa}^{[\sa]}$ so 
(a)-isoscelesness and collinearity are nodal, with (a)-regularness elsewise favoured,
Finally, the R = 2 = r sin orbital is $d_{\sd\sr\sa_x\sd\sr\sa_y}^{[\sa]} = d_{\sA\sn\si\sss\so
\,\sE\sll\sll\si\sp}^{[\sa]}$ so (a)-isoscelesness and (a)-regularness are nodal, with collinearity 
elsewise favoured, and the cos orbital is $d_{\sd\sr\sa_x^2 - \sd\sr\sa_y^2}^{[\sa]} = 
d_{\sA\sn\si\sss\so^2 - \sE\sll\sll\si\sp^2}^{[\sa]}$ so the (a)-isosceles and (a)-regular parts of the 
collinearity plane are favoured.

I also use the (a)-bases extension of \cite{08II}'s (1)-basis. 
These bases are important through each being associated with one pure relative angular momentum quantity 
${\cal J}_{(\sa)}$, so the projected quantum number r is here a pure relative angular angular momentum 
quantum number $\mj_{(\sa)}$, and through alignment with the remaining axis of symmetry upon switching 
on a special HO potential.
In these bases, the form of the solution is
\beq
\Psi_{\sR\sj_{(\ta)}}(\Theta_{(\sa)}, \Phi_{(\sa)}) \propto 
\mP_{\sR}^{\sj_{(\ta)}}(\mbox{Ellip}(\ma)){\cal T}_{\sj_{(\ta)}}
\left(
\mbox{Aniso(a)}/\sqrt{1 - \mbox{Ellip(a)}^2}
\right) \mbox{ }    
\label{terfu}
\eeq
(which is a rather tidier form than in \cite{08II} through the identification of shape quantities and use of the 
${\cal T}$ symbol.)
The physical meaning of the triangleland orbitals in this basis is then as follows.   
One has the same $s$-orbital as usual, while the $p^{(\sa)}$, R = 2, j$_{(\sa)}$ = 1 and R = 2, 
$j_{(\sa)} = 2$ cos solutions are among the previous paragraph's orbitals under exchange of y's and z's.  
The R = 2, j$_{(\sa)}$ = 0 solution has orbital $d^{(\sa)}_{\sd\sr\sa_z\mbox{}^2} = 
d^{(\sa)}_{\sE\sll\sll\si\sp^2}$ so it has a ring of (a)-regularness and lobes of (a)-sharp and (a)-flat 
triangles.  
Finally, the R = 2, j$_{(\sa)}$ = 2  sine solution has orbital $d^{(\sa)}_{\sd\sr\sa_x\mbox{}^2 - 
\sd\sr\sa_y\mbox{}^2} = d^{(\sa)}_{\sA\sn\si\sss\so^2 - \sT\se\st\sr\sa\sA\sr\se\sa^2}$ so favouring 
collinear and (a)-isosceles (i.e. equilateral) parts of the plane of (a)-regularness with nodes in 
between.

As examples of results that would be very complicated in \cite{08II}'s presentation, 1) in the 
$(\gamma)$-bases (for which the projected quantum number is a distinct $r_{(\gamma)}$ for each 
$\gamma$),  

\noindent
\beq
\Psi_{\sR\sr_{(\gamma)}} \propto 
\mP_{\sR}^{r_{(\gamma)}}(\mbox{Ellip(1) cos}\,\gamma + \mbox{Aniso(1) sin}\,\gamma)
{\cal T}_{\sr_{(\gamma)}}
\left( 
\frac{\mbox{Aniso(1) sin}\,\gamma - \mbox{Ellip(1) cos}\,\gamma }
{\sqrt{1 - (\mbox{Aniso(1)}\,\mbox{sin}\,\gamma\, + \mbox{Ellip(1)} \, \mbox{cos}\,\gamma )^2}}
\right) \mbox{ }   
\eeq
2) In the $[\gamma]$-bases, using for $\overline{\cal T}_{\sr}$ of the cos solution  = ${\cal T}_{\sr}$ 
of the sin one and $\overline{\cal T}_{\sr}$ of the sin solution  = $-{\cal T}_{\sr}$ of the cos one, 
\beq
\Psi_{\sR\sr} \propto \mP_{\sR}^{\sr}(\mbox{TetraArea})
\left[
         {\cal T}_{\sr}\left(\mbox{Ellip(1)}/\sqrt{1 - \mbox{TetraArea}^2}\right)\mbox{cos}\,r\gamma +
\overline{\cal T}_{\sr}\left(\mbox{Ellip(1)}/\sqrt{1 - \mbox{TetraArea}^2}\right)\mbox{sin}\,r\gamma
\right] \mbox{ } .     
\eeq

\subsection{QM with nontrivial potentials on triangleland}

Near the \{23\} double collision, to second approximation, this problem gives \cite{08II} 2-$d$  isotropic 
HO wavefunctions \cite{Messiah, SchwRo} (but not the same inner product in 
detail, due to the curved geometry).
I work in the ${(1)}$-basis that is adapted to this potential, but suppress the (1) labels, and I 
require a quantum-mechanically sizeable classical `frequency' (dimensionally frequency 
$\times \mbox{ } I$ in this paper's formulation) $\omega$ so the bulk of the wavefunction lies where the 
small-angle approximation holds (e.g. $\omega/\hbar$ of the order of $10^3$).
Then the `energies' are
$
\fE - {K_2}/{2} = \mn\hbar\omega 
\mbox{ } \mbox{ for } \mbox{ } 
\mn := 1 + 2\mN + |\mj| 
$,
though, $\omega$ itself depends on the shifted energy, $\fE^{\prime} := \fE - A - B$, so 
$\fE^{\prime} = 2
\big(
\hbar^2\mn^2 + \mn\hbar\sqrt{\mn^2\hbar^2  - B}
\big),$ 
which, for $\mn\hbar/\omega$ small i.e. $\mn << 10^3$ (which is certainly OK for the 
solutions below), goes as $\fE^{\prime} \approx \mn\hbar\Omega$ for $\Omega = 2\sqrt{-B}$.  
The solutions go like 
\beq
\Psi_{\sN\sj}(\Theta, \Phi) \propto {\Theta}\mbox{}^{|\sj|}
\left(
1 + |\mj|\Theta\mbox{}^2/{12}
\right)
\mbox{exp}
\left(
- \omega\Theta\mbox{}^2/{8\hbar}
\right)
\mL_{\sN}^{|\sj|}
\left(
{\omega\Theta\mbox{}^2}/{4\hbar}
\right)
\mbox{exp}(\pm i\mj\Phi)
\label{gi}
\eeq
where $\mL_{a}^b(\xi)$ is the associated Laguerre polynomials in $\xi$.  
[The $\Phi$ factor of this is rewriteable in terms of Aniso and Ellip as in (\ref{terfu}), while the 
$\Theta$ factor is now a somewhat more complicated function of Ellip.]

Interpreting the solutions, 
again, r = j = 0 are axisymmetric and as such are totally undiscerning of whichever notion of 
isoscelesness or regularity.    
N = 0, j = 0 favours sharp triangles.  
N = 0, j = $\pm 1$ are a degenerate pair, the sin and cos solutions of which favour sharp triangles that 
are approximately collinear and isosceles respectively.  
The next 3 solutions are also degenerate. 
N = 1, j = 0 favours two separate bands: the fairly sharp triangles and the even sharper triangles.  
The N = 0, $|\mj| = 2$ cos solution favours sharp triangles that are approximately collinear or 
isosceles, while the sin solution favours sharp triangles that are neither. 
The N = 0 states have peaks at $\Theta \approx 2\sqrt{{\cal J}/\omega}$; c.f. the means in the next Sec.  
For the opposite sign of $B$, the other pole's [(1)-merger] approximate wavefunctions have the same 
interpretation except that one replaces `sharp' by `flat'.

\section{Interpretation by use of shape operators}

Use of these is inspired by a parallel with atomic physics, as outlined in \cite{AF}, which makes use of 
`Wigner 3j symbol' type mathematics \cite{Mizu}. 
I thus obtain that, in the (a) basis for the free problem, 
$
\langle\mR\,\mj_{(\sa)}\,|\,\widehat{\mbox{Aniso(a)}} \,|\,\mR\,\mj_{(\sa)}\rangle = 
\langle\mR\,\mj_{(\sa)}\,|\,\widehat{\mbox{TetraArea}}\,|\,\mR\,\mj_{(\sa)}\rangle = 
\langle\mR\,\mj_{(\sa)}\,|\,\widehat{\mbox{Ellip(a)}} \,|\,\mR\,\mj_{(\sa)}\rangle = 0  
$, signifying that there is orientation symmetry so each positive contribution is cancelled by a 
corresponding negative one.  
The useful information starts with the spreads, 
\beq
\Delta_{\sR\,\sj_{(\ta)}}(\widehat{\mbox{Ellip(a)}}) = 
\sqrt{\frac{2(\mR(\mR + 1) - {\mj_{(\sa)}}^2) - 1}
           {(2\mR - 1)(2\mR + 3)}} \mbox{ } , 
\label{110}
\eeq
\beq
\Delta_{\sR\,\sj_{(\ta)}}(\widehat{\mbox{TetraArea}}) = 
\sqrt{\frac{(\mR(\mR + 1) + {\mj_{(\sa)}}^2) - 1}
             {(2\mR - 1)(2\mR + 3)}                Q_{2}(\mj_{(\sa)})} 
\mbox{ } , \mbox{ }
\Delta_{\sR\,\sj_{(\ta)}}(\widehat{\mbox{Aniso(a)}}) = 
\sqrt{\frac{(\mR(\mR + 1) + {\mj_{(\sa)}}^2) - 1}
           {(2\mR - 1)(2\mR + 3)}                  Q_{1}(\mj_{(\sa)})} \mbox{ } ,  
\label{310}
\eeq
for 
$Q_2(\mj_{(\sa)}) =      1/2 \mbox{ for } \mj_{(\sa)} \mbox{ cosine solution, } \mbox{ }  
                         3/2 \mbox{ for } \mj_{(\sa)} \mbox{ sine solution, } \mbox{ }
                         1   \mbox{ otherwise}$, 
and $Q_1(\mj_{(\sa)})$ the sin $\leftrightarrow$ cos of this.   
Thus, the ground state spreads in Ellip(a) and Aniso(a) are $\frac{1}{\sqrt{3}}$. 
Also, the spread in $\widehat{\mbox{Ellip(a)}}$ goes as $\frac{1}{\sqrt{2\sR}}$ for j$_{(\sa)}$ maximal 
and R large, and as $\frac{1}{\sqrt{2}}$ for j$_{(\sa)}$ = 0 and R large.  
The former amounts to recovery of the equatorial classical geodesic as the limit of an ever-thinner belt 
in the limit of large maximal relative angular momentum quantum number $|\mj_{(\sa)}| = \mR$ (traversed 
in either direction according to the sign of j$_{(\sa)}$). 
The latter amounts to the $s$, $p^{(\sa)}_{\sd\sr\sa_z}$, $d^{(\sa)}_{\sd\sr\sa_z\mbox{}^2}$ ... 
sequence of orbitals not getting much narrower as R increases, so that there is always limited 
concentration of probability on sharp or flat triangles.  
$\Delta(\widehat{\mbox{TetraArea}})$ goes as $\frac{1}{\sqrt{2}}$ for large quantum number with 
j$_{(\sa)}$ maximal and as $\frac{1}{2}$ for large quantum number with j$_{(\sa)}$ = 0.  
That the wavefunctions have increasingly many peaks and valleys does not register unto this overall 
spread quantifier.  
Also, one can see that there are no limiting collinear states in this basis.  
This $\frac{1}{\sqrt{2}}$ is the maximal value, while the minimal one, $\frac{1}{\sqrt{5}}$, is for 
R = 1, j = 0.  
For the ground state, $\Delta_{0\,0} (\widehat{\mbox{TetraArea}})$ = $\frac{1}{\sqrt{3}}$.

Next, in the equilateral basis, the free problem gives the TetraArea $\leftrightarrow$ Ellip(a) and 
$\mj_{(\sa)} \rightarrow \mr$ counterpart of the preceding paragraph's equations.    
Thus one does now get a collinear limit for large quantum numbers, e.g. in the case of r maximal, 
$\Delta_{\sR\,\sr}(\widehat{\mbox{TetraArea}}) \approx \frac{1}{\sqrt{2\sR}} \longrightarrow 0$ for r 
maximal and R large. 
On the other hand, this tends to  $\frac{1}{\sqrt{2}}$ for r = 0 and R large.

Finally, in the small regime for the HO problem, TetraArea and Aniso continue to have zero expectation.  
For N, $\mj << \frac{\omega}{\hbar}$ 
%
$\langle \mN\,\mj\,|\,\widehat{\mbox{Ellip}}\,|\,\mN\,\mj\rangle \approx 
1 - \frac{\sn\hbar}{2\omega}$, 
$\Delta_{\sN\,\sj}(\widehat{\mbox{TetraArea}}) \approx 
\sqrt{         \frac{\sn\hbar Q_2(\sj)}{2\omega}       }$, 
$\Delta_{\sN\,\sj}(\widehat{\mbox{Aniso}})   \approx 
\sqrt{        \frac{{\sn}\hbar Q_1(\sj)}{2\omega} }$ 
Also, the spread in Ellip goes as $\frac{\hbar}{\omega}$ though the scheme cannot really evaluate its 
coefficient (since there is cancellation of the orders to which the approximations made hold).
This gives some idea of what unsigned area is typical, with $\Delta_{00}(\widehat{\mbox{TetraArea}})$ 
having an interpretation somewhat akin to the Bohr radius as a `smallest amount'. 
The change in spread of TetraArea due to the potential's confining effect is from a wide range of 
areas corresponding to a spread of $\frac{1}{\sqrt{3}}$ 
to $\sqrt{\frac{\hbar}{2\omega}}$ for $\frac{\omega}{\hbar}$ large, i.e. smaller by a factor of 
$\sqrt{\frac{3\hbar}{2\omega}}$.

\section{Na\"{\i}ve Schr\"{o}dinger Interpretation}

The na\"{\i}ve Schr\"{o}dinger interpretation \cite{NSI} -- a simple timeless approach to the Problem 
of Time which gives probabilities of the universe possessing some particular property (with no 
reference to when or to dynamical evolution/history).      
I make use of the configuration space regions of Fig 2d) that correspond to various physically 
meaningful criteria on the model universe triangles. 
E.g. 1) Dropping [1]-basis subscripts, 
$
\mbox{P(triangular model universe is $\epsilon$-equilateral)} \propto  
\int_{\epsilon\mbox{\scriptsize -caps}}|\Psi|^2 \d S = \int_{\Phi = 0}^{2\pi}
\left(
\int_{\Theta = 0}^{\epsilon} + \int_{\pi - \epsilon}^{\pi}
\right)
|\Psi(\Theta, \Phi)|^2 \mbox{sin}\,\Theta\,\d \Theta\d\Phi$, which, in the very special case, is 
$\propto$ $\epsilon^2 + O(\epsilon^4)$ for the ground state and R = 1, r = 0, and $\propto$ 
$\epsilon^4$ + $O(\epsilon^6)$ for R = 1 = $|$r$|$.  
\noindent
E.g. 2) Corresponding to the magnitude of TetraArea not exceeding the small number $\delta$, 
P(triangular model universe is $\delta$-collinear) $\propto \int_{\delta\mbox{\scriptsize -belt}}
|\Psi|^2 \d S$, which, in the very special case is $\propto$ $\delta + O(\delta^3)$ for the ground state 
and R = 1 = $|$r$|$, and $\propto \delta^3$ for R = 1, r = 0.  
1) conforms with R = 1, $|$r$|$ = 1 pointing along axes in the plane of collinearity, and 
2) with R = 1, r = 0 pointing along the E-axis with a node in the plane of collinearity.
Also, 2) continues to make sense for the small regime of the special problem.  
The result here for the lowest 4 states goes as
$\mbox{P(triangular model is $\delta$-collinear)} \propto \delta\sqrt{\omega/\hbar}$,so that there is a 
sizeable concentrating factor $\sqrt{\omega/\hbar}$ as compared to the very special case, i.e. the 
potential is trapping more of the wavefunction near the collinearity plane.

\section{Conclusion}

In this paper, I applied methods analogous to \cite{AF}'s for the 4-stop metroland RPM to improve 
\cite{08I, 08II}'s understanding of the triangleland RPM (both of these RPM's having $\mathbb{S}^2$ 
shape spaces, albeit with big differences in the physical interpretation of each).    
Two of these methods are

\noindent 
A) Tessellation of the shape sphere by the corresponding physical interpretation. 
For triangleland, this has double collision points, equilateral points, an equator of collinearity and 
meridians of isoscelesness and rightness.
This tessellation can then be used as a back-cloth to read off which types of triangles are picked out 
by classical trajectories and by potentials and wavefunctions as height functions over the shape sphere. 

\noindent
B) The presentation of the sphere in terms of 3 quantities u$^{\Delta}$ the sum of whose squares is 1. 
For 4-stop metroland, they are just the obvious relative separations playing the role of unit 
Cartesian vectors for the $\mathbb{R}^3$ of relative coordinates that surrounds the shape sphere. 
However for triangleland they are a less trivial Hopf/Dragt realization.  
I furthermore interpret the ensuing three Dragt-type coordinates as the following shape quantities. 
I) Aniso, a measure of anisoscelesness with respect to a given clustering that assigns which particles 
are to be the apex and base). 
II) Ellip, the ellipticity of the moments of inertia with respect to a given clustering. 
III) TetraArea, a quantity proportional to the area of the triangle, and which is in some senses deeper 
than the preceding though being clustering-independent alias a `democratic invariant', of which 4-stop 
metroland has no counterpart.\footnote{The tessellation technique and these nice geometric 
interpretations for Dragt-type coordinates are partly new to this paper, and should also be of interest 
in Molecular Physics (the subject Dragt coordinates originate from; also N.B. there are a number of 
distinctions between the conventional 3-particle molecule and this paper's model, due to the latter 
being a model for whole-universe quantum cosmological use.)}
%
I promote these quantities at the quantum level to shape operators, which provides an `expectations and 
spreads' account of the wavefunctions that complements \cite{08I} and Sec 6's `modes and nodes' account.  
Moreover, I use not only \cite{08II}'s orbital basis centred about a particular clustering, and 
permutations thereof, but also the `natural bases' centred about the equilateral configurations, which 
are more natural in the absense of nontrivial potentials.  
I also consider the effect of a confining potential and provide a superior account of the nature of 
triangleland's conserved quantities.  
N.B. the mathematical analogy with the 2-sphere in space analogue makes it clear that the 
{\sl mathematics} underlying this paper is of a simple form that is very standard in QM.  
How this mathematics arises as a simple example of the more general machinery for quantizing quotient 
spaces is covered e.g. in \cite{Isham84}.
What is different from the usual in this paper is the {\sl subsequent interpretation} of this standard 
mathematics in the Hopf/Dragt-type way that is appropriate for triangleland.
Moreover, this interpretation renders this simple (and thus useable with all its additional well-known 
machinery) mathematics so as to address, compare and combine further Problem of Time in Quantum Gravity 
strategies and quantum cosmological workings.

Pure-shape RPM's such as this paper's are more straightforward in a number of ways than RPM's that 
contain scale as well. 
However, it is scaled RPM's are more tightly analogous to standard Classical and Quantum Cosmology
(and hidden time and emergent semiclassical time strategies for the Problem of Time), and so these are 
also under study \cite{MGM, Cones, ScaleQM, 08III}.  
It is thus important to note that pure-shape RPM's occurs within scaled ones as a subproblem, so 
it makes good sense to study the pure-shape case first, with the present paper's results retaining 
their relevance for scaled RPM's.

The free solutions studied in this paper and \cite{08II, AF} have the further value of supporting a 
technically straightforward (Wigner 3j symbol \cite{Mizu}) perturbation theory thereabout, allowing for 
treatment of more complicated potential terms, provided that they are small. 
In the emergent semiclassical approach setting, scaled triangleland \cite{08III} is a toy model of 

\noindent 
the Halliwell--Hawking approach to the study of the quantum cosmological origin of small inhomogeneity 
(shape) perturbations in GR cosmology.  
This paper's tessellation is useful for characterization of regions in which various semiclassical 
approaches work \cite{08III, Forth}. 
Triangleland is also a model for the internal time \cite{06I, SemiclI}, histories theory \cite{08III, 
Forth} and timeless records theory \cite{Records, Forth} approaches to the Problem of Time (this 
paper's na\"{\i}ve Schr\"{o}dinger Inerpretation is a simpler timeless approach).
Triangleland is, furthermore, simple enough that many of these will be solvable, comparable and 
combineable (e.g. \cite{EOT, H99, H03, H09} express great interest in such combinations).  
Some such applications may benefit from extending the modelling from 4-stop metroland and triangleland 
to quadrilateralland (for which \cite{QShape} is the counterpart of the present paper as regards shape 
quantities).
I end with 2 quantum cosmological applications that this paper makes contact with.  

\noindent  
1) A further difference between this paper's models and (at least many of) their Molecular Physics 
analogues is at the level of which `multiverses' correspond to each. 
(For Molecular Physics models the corresponding multiverse can be thought of as a collection of 
laboratory setups with different parameter values; multiverses in the RPM context are at least 
aesthetically more appealing in distancing themselves from Copenhagen interpretation connotations.) 
In particular, the triangleland HO problem has many mathematical parallels with the Stark effect for 
the linear rigid rotor \cite{08II}.
However, in the latter, the electric field can point in any 3-$d$ direction, while in the former, the 
corresponding direction favoured by the potential cannot be rotated out of the collinearity plane.  
Moreover, the triangleland version has not 1 but 3 DM axes of particular physical significance within 
this plane (Fig 2b).   
Thus each of these problems is set in a different `multiverse'.  

\noindent
2) Whether uniform states are probable (see e.g. \cite{Penrose}) is of considerable interest in Quantum 
Cosmology -- and Cosmology in general -- due to our actual universe being fairly uniform and expected 
to have emerged from an extremely uniform state.  
Triangleland's most uniform states are equilateral triangles; the natural basis new to this paper is 
then useful for studying propositions about these; $\epsilon$-caps about the poles (Fig 2d) in this 
basis represent approximately uniform states -- ``$\epsilon$-equilateral states.      
Additionally, (approximate) collinearity can be seen as the opposite of this [and both of these are 
quantifiable in terms of TetraArea, which is minimal in size for collinearity and maximal for the 
equilateral configurations: $\epsilon$-equilateral corresponds to TetraArea lying between 1 and 1 -- 
$\epsilon^2/2$ for $\epsilon$ small ($<< 1$)].  
%
%
For this paper's very special HO case, going through the $s, p_{u_z}, d_{u_z^2} ...$ sequence of 
wavefunction sequence leads to only a limited increase in peaking around the equilateral configurations 
so that the most uniform states -- the two labellings of the equilateral triangle are never 
overwelmingly the most probable states.
Furthermore, switching on the special or general HO potential draws the probability density away into a 
well that is not aligned with the equilateral configurations but rather within the collinearity plane.  
This paper further investigated these most and least uniform states via the Na\"{\i}ve Schr\"{o}dinger 
Interpretation, by which I found that highly uniform states are not especially favoured, whilst 
the presence of a simple HO-like potential increases the probability of its opposite, near-collinearity.

\mbox{ } 

\noindent{\bf Acknowledgements}  
My wife Claire for her support. 
Miss Anne Franzen, Prof Don Page, Prof Jonathan Halliwell, Prof Gary Gibbons, Visiting Prof Julian 
Barbour and the anonymous referees for comments, discussions and references, and the Bad Honnef 
100th Anniversary of Minkowski Spacetime Conference for inviting me to speak.


\end{document}